\documentclass[11pt,preprint]{aastex}

\newcommand\hb{H${\beta}$~}
\newcommand\ha{H${\alpha}$~}

\def\nii{[N~{\sc ii}]}

\def\heii{He~{\sc ii}}
\def\oiii{[O~{\sc iii}]}
\def\hi{H~{\sc i}}
  
\slugcomment{}
\shortauthors{Stanghellini et al.}
\shorttitle{Planetary Nebula Morphology and Evolution} 

\begin{document}

\title{The Correlations between Planetary Nebula Morphology and Central Star
Evolution. Analysis of the Northern Galactic Sample.}

\author{Letizia Stanghellini\altaffilmark{1,2}}
\author{Eva Villaver\altaffilmark{1}}
\author{Arturo Manchado\altaffilmark{3,4}}
\author{Martin A. Guerrero\altaffilmark{5}}

\altaffiltext{1}{Space Telescope Science Institute, 3700 San Martin Drive,
Baltimore MD 21218, USA; lstanghe@stsci.edu; villaver@stsci.edu}
\altaffiltext{2}{Affiliated to the Astrophysics Division, Space Science
Department of ESA; on leave, INAF- Osservatorio Astronomico di Bologna}
\altaffiltext{3}{Instituto de Astrofisica de Canarias, via Lactea s/n, 
La Laguna, E-38200 Santa Cruz de Tenerife, Spain; amt@iac.es}
\altaffiltext{4}{Consejo Superior de Investigaciones Cientificas, Spain}
\altaffiltext{5}{Department of Astronomy, University of Illinois at
Urbana-Champaign, 1002 W. Green St., Urbana, IL 61801; mar@uiuc.edu}
\newpage

\begin{abstract}
Northern Galactic Planetary nebulae (PNs) are studied to 
disclose possible correlations between the morphology of the nebulae
and the evolution of the central stars (CSs).
To this end, we have built the best database available to date, 
accounting for homogeneity and completeness. We use updated statistical
distances, an updated morphological classification scheme, and 
we calculate Zanstra temperatures for a large sample of PNs.
With our study
we confirm that round, elliptical, and bipolar
PNs have different spatial distributions within the
Galaxy, with average absolute distances to the Galactic plane 
0.73, 0.38, and 0.21 kpc respectively.
We also find evidence that the distributions of the
central star's masses are different across these
morphological groups, although we do not find that CSs
hosted by bipolar PNs are hotter, on average, than CSs
within round and elliptical PNs.
Our results are in broad agreement with the
previous analyses, indicating that round, elliptical, and bipolar
PNs evolve from progenitors in different mass ranges, and may
belong to different stellar populations, as also indicated by the 
helium and nitrogen abundances of PNs of different morphology. 
\end{abstract}

\keywords{Planetary Nebulae: Morphology, Evolution. Central stars:
Progenitors, Evolution}

\section{Introduction} 

Stars with masses in the 1--8 M$_{\odot}$
range are very likely to go through the evolutionary phase of PN
ejection at the end of their lives. The ejection of most 
of the gaseous envelope in Asymptotic Giant Branch (AGB) stars
occurs at the tip of the Thermal Pulse phase, in the form of a 
massive, low velocity wind \citep{ir83}.
Subsequently,
the remnant CS ionizes the gaseous ejecta, while
the fast, low mass-loss rate CS wind shapes the PN. The observed
PN morphology thus depends on a complicate combination of phenomena, 
some occurring within the nebular gas, which evolves in dynamic time scale, 
others
deriving from the evolution of the stellar progenitors and of the CSs.
Morphology may also depend on the physical status of the
interstellar environment of the PN progenitor.

The relations between PN morphology and CS
evolution may be used to explore the gas versus star interplay.
Pioneering this field, 
\citet{pei78} found a link between PN 
morphology and stellar evolution through the chemical elements that are produced
by PN progenitors, and are drawn to the stellar surface via dredge-up
episodes. Peimbert's study showed that most of the helium- and nitrogen-rich, carbon-poor
PNs where of asymmetric (bipolar etc.) shapes. 
This result is broadly consistent with the predictions 
of stellar evolution if the progenitors 
of asymmetric PNs have on average larger masses than the progenitors of 
round (and elliptical) PNs, without any assumption or relation to stellar
multiplicity of the stellar progenitors. 

About a decade ago, several authors have tried to tie up the correlations
between nebular morphology and stellar properties \citep{scs93,cs95}, 
inspired by the
new bounty of narrowband images available for Galactic PNs
\citep{scm92,bal87}. \citet{scs93}, based 
on a large, yet non homogeneously selected, sample of Southern PNs from
the ESO catalog \citep{scm92}, 
found that the
CSs of symmetric and asymmetric PNs have different mass 
distribution, and that the CSs of bipolar PNs are spread
over a large mass range.
Other papers have been written in the recent years \citep{amn95,gst97}
on this subject, none of which have challenged the earlier results. 

Strength to the association of bipolar PNs with high mass progenitors comes 
from the Galactic distribution of morphological types. Among
others, \citet{scsh93}, \citet{sta95}, and \citet{mea00} found that bipolar PNs
lie closer to the Galactic plane than round and elliptical PNs,
making the high-mass origin plausible.

The motivation of the present work is to test the correlations 
between PN morphology and CS evolution with a larger and homogeneous
PN database.
In 1996 the IAC Catalog of PNs \citep{mea96}, and with it a complete and
homogeneous PN image data set, became available. 

The aim of this paper is to discuss the relations between morphological 
type and the physical  parameters that are linked to stellar populations
and evolution. We will examine principally the variations of 
temperature, luminosity, and mass of the CSs. We will also
show the variation across morphological types of elemental abundances
and Galactic distribution.
The statistics of the
nebular characteristics variations across morphological 
types will be published elsewhere \citep{mea02}.

In $\S$2 we discuss the sample selection and the morphological 
classification. In $\S$ 3 we describe the determination of the
stellar temperature and luminosity.
In $\S$ 4 we present the results of our analysis. A discussion 
of the significance of our findings for the understanding of 
the evolution of PNs and their CSs is presented in $\S$5.

\section{Morphological classification}

The database that we have used in this paper includes the IAC 
morphological catalog of
Northern Galactic PNs \citep{mea96}, supplemented by 
a selection of PNs from \citet{bal87} and \citet{scm92}. 
Therefore, our sample includes all PNs in 
\citet{aea92} that are
larger than 4 arcsec in maximum size, and that are within the observing limits 
of the IAC and La Palma Observatory telescopes \citep{mea00}.

We classify the PN
morphology primary on the basis of the \ha images, and secondary on 
the \oiii~ 5007 and the \nii~ 6584 images, when available. A thorough
discussion on the extended morphological classification
will be available in \citet{mea02}.
Here, we briefly describe the morphological classes used to analyze
the results on the CSs correlations.

We use the following, simplified morphological scheme:
\begin{itemize}
\item{Round (R) PNs.}
\item{Elliptical (E) PNs. We distinguish between R and E PNs if 
an axial difference of at least 5 $\%$ is detected.}
\item{Bipolar (B) PNs. As commonly defined, bipolarity implies 
at least one pair of lobes and 
a pinched {\it waist}. Quadrupolar
PNs belong to this category.}
\end{itemize}.

Together with the three main morphological classes above, we also examine
the group properties of 
Bipolar Core (BC) PNs. This morphological class identifies those PNs whose external contour
does not show obvious bipolar lobes, but whose inner structure is
clearly suggestive of a projected ring. A ring-like structure was detected
above the 20 $\%$ flux level for our BC PNs to be classified as such. This class was firstly
used in \citet{sea99} to analyze bipolarity 
in Magellanic Cloud PN images acquired before the first HST servicing
mission that restored the telescope optics. Note that \citet{gst97}
use a similar definition for embedded bipolars (BE).

A few comments on this simplified classification scheme are in order. 
First note
that \citet{scs93} do not distinguish between R and E PNs.
We now know that it is possible to separate these classes, and
that they have different properties.
Note that our classification of round PN is an objective one, based on the
measured axial ratio on the 2D images, without making
assuptions
on the 3D structures (note that \citet{sok2002} 
uses a different classsification 
approach, and subjectively defines elliptical those PNs that have structures
such as knots or filaments, even if they have round contours). 

Second, in this paper we subsume the quadrupolar PNs into the bipolar 
class. 
Quadrupolar (Q) PNs show two pairs of lobes \citep{msg96} and their formation 
mechanism has to be similar to that of bipolar PNs. Our database 
contains but a few Q PNs, and studying them as a class
apart will dilute their statistical significance anyway.

Third, we do not classify pointsymmetric (P) PNs as a separate 
morphological type.
As pointed out by \citet{mea00}, pointsymmetric
structures are present in a variety of main body morphologies. For this
reason, we classified the pointsymmetric PNs as round, bipolar, or
elliptical depending on the structure of the main body of the PN.
Pointsymmetry is detected in a very small fraction of the PNs 
in our sample, and if they were selected in a separate class the 
results from our analysis would not change statistically.
Finally, it is worth noting that
the bipolar core PNs would have been classified as R or E 
in the usual or the new morphological schemes \citep{mea02}. We 
want to isolate their properties in this paper, and relate them to those of 
B PNs. By doing so, we moderately decrease the R and E PN data sets.

\section{Temperature and luminosity derivation}

The stellar properties of the selected PNs were 
determined via the Zanstra analysis \citep{kal83}. 
The input parameters to the Zanstra analysis were selected
from the literature, with the endeavor of using the 
best possible set of input data, chosen as homogeneously as possible.
In Table 1 we list our reference choices for the input parameters. 

The stellar magnitudes used for the analysis were taken, for the 
most part, from the paper by \citet{tea91}.

The \hb and \heii~($\lambda$4686) line intensities were drawn, when available,
from \citet{cks92}, otherwise from \citet{tea91}. Additional fluxes
and extinction constant references are listed in Table 1. 
Extinction constants are from \citet{cks92} or, alternatively,
from \citet{sea92}. If neither of these two references had relevant
data, we used the additional references listed in table 1.

Nebular radii were measured by us on the original frames.
For elliptical PNs, we compute equivalent radii as
$R_{\rm E}=[{(a \times b)/4}] ^ {1/2}$, where $a$ and $b$ are the elliptical 
axes. For bipolar PNs, we make the approximation that 
the lobes are two ellipses, and we define 
$R_{\rm E}=[{(a \times b)/4}] ^ {1/2}$, where $a$ is the extension
of the waist, and $b$ the extension of the bipolar lobes.

All distances used in this paper are statistical distances. We have
calculated all the statistical distances based on the newly measured 
radii, 
and the \citet{cks92} distance scale calibration.
The 6~cm fluxes used in this
calculation were taken from \citet{cks92} where available, otherwise from
the other references in Table 1.
To assure that the calibration of the distance scale is independent 
on the morphological type of the calibrators used, we have morphologically
classified the calibrators in \citet{cks92}, and verified that all 
morphological types have similar weight in the calibration. We also
verified that the calibration based on just bipolar and bipolar core PNs
is almost identical to the calibration based on round PNs.

All input data are used in our Zanstra code with their own original error bars, 
as given from the
original papers listed above and in Table 1. The Zanstra analysis then produces the 
propagated error bars. 
The statistical distances do not carry error bars, as it is impossible to
quantify the magnitude of the errors in the assumptions made to scale
the ionized masses with the nebular distances (see the discussion by \citet{cks92}).

\section{Results}

In the following sections we illustrate the results of our investigation.
Among the 255 Northern Galactic PNs with morphological classification, 
Zanstra analysis was performed in all the cases where the input
parameters were available in the literature, and precisely for 54 round, 96
elliptical, 36 bipolar, and 19 bipolar core PNs. 

In Table 2 we give the results of the Zanstra analysis. Column (1) gives the 
PN name; column (2) gives the morphological type, following the simplified
scheme described in $\S$2. Unless otherwise noted within the Table
the images for morphological analysis were from \citet{mea96}. 
Columns (3) through (6) give the Zanstra temperature (and error) and
luminosity (and error), in logarithmic format. We always give the \heii~
Zanstra temperature if available. Otherwise, we give the \hi~ Zanstra 
temperature. In the cases in which the stellar magnitude was not available,
we give the cross-over temperature instead \citep{kj89}. 
The reader should not use cross-over temperatures in Table 2 
at face value, since they
are calculated assuming that the \heii~ and \hi~ temperatures are 
equal. Cross-over temperatures are upper limits for most nebulae except for 
optically thick PNs, and the magnitudes estimated via cross-over analysis are
also upper limits. 

Distances, dimensions, and detailed morphology will 
be discussed in a future paper.

In Table 3 and 4 we give the averaged physical diagnostics across morphological types
that we use in the following sections.
Table 3 contains the Galactic distribution and Zanstra temperature diagnostics (with
the $\sigma$ values in square braketts),
while Table 4 gives the elemental abundance diagnostics.

\subsection{Spatial distribution}

The apparent spatial distribution of Galactic PNs with different 
morphology is shown in 
Figure 1, where  the different symbols indicate morphological types 
(open circles are round, asterisks are elliptical, filled squares
are bipolar, and open squares are bipolar core PNs).
It appears from this plot that the bipolar
and bipolar core PNs are found at lower Galactic latitudes, and their
distribution do not extend, for the most part, away from the disk. Differently,
elliptical and round PNs are spread all
over the l-b plane. From column (2) in Table 3 we can see that the round PNs
have a latitude range much larger than that of bipolar PNs.

We have run the Kolmogorov-Smirnov (K-S) statistics \citep{pea86} on 
the Galactic latitude distribution to investigate the quantitative 
differences of the different morphology PNs. 
The K-S statistical test give the maximum absolute value of the difference
between two given distributions, D$_{\rm max}$. The significance level
for the null hypothesis that the data sets are drawn from the same distribution 
(P) is also given by the K-S statistics: lower values of P indicate two
different distributions. Typically,  P $\le 5\times10^{-2}$ indicates that
there is a real possibility 
that the two distributions are different, while P=1 indicates identical
distributions. Larger data sets give better
results, but data sets as large as N=20 already give a good indication on the
statistics. 

By considering 
the Galactic latitude distributions of R and B PNs, we found that
D$_{\rm max}$=1.8 and P=0.003. The latitude distributions of 
round and elliptical PNs are not so sharply dissimilar (P=0.19). 
To enhance the statistical significance of the results we 
grouped 
B and BC PNs in one class (B+BC). The K-S probability 
for R and B+BC PNs to have similar {\it b} 
distributions is 
0.086, while this probabilities is 0.53 when
we use the E and B+BC samples. Our conclusion is that there is a 
sharp difference between the 
nature of R and B+BC PNs. Much more detail on these distribution 
will be presented in a future paper. 

In Figure 2 we show the PNs distribution of the absolute distance to the 
Galactic plane, $|z|$, for each morphological type. This
parameter is calculated with the statistical distance scale, 
and the sample sizes are different than those of Figure 1, since
some PNs did not have the necessary parameters in the literature for
the derivation of their statistical distance. We see in Figure 2
that the distribution of R PNs is the broadest, followed by the E,
the BC, and the B PNs. 

We have calculated the average scale distance from the Galactic plane
for the different morphological types (column [3] in Table 3). 
Round PNs have the largest average $|z|$, E PNs follow,
then BC PNs , and finally B PNs, closest to
the Galactic plane. The average scale height 
of all PNs in the sample is 0.44 kpc. 
Our calculations basically confirm the early results by \citet{scsh93}.
It is worth noting that the average distance per morphological class
are comparable to one another.

Our results on the Galactic distribution of morphological types 
show that morphology could be used as indicator of 
stellar population. For example, the scale height for a young
disk population is similar to that of bipolar and bipolar core PNs
\citep{sca98}.
The caveat of this conclusion is that there may
be a selection effect that limits the observations of
low surface brightness PNs toward the Galactic plane.

In order to asses the importance of the selection effects,
we checked the \hb surface brightness distributions of the PNs, 
calculated as F$_{\beta} / \pi R^2$, where the \hb flux is corrected
for extinction and the radii are the equivalent radii calculated
as described in $\S$3.
We found that surface brightness of R and E PNs
are in the $-17 < {\rm log}~ SB_{\beta} < -11.5$ range, while
B and BC PN's surface brightness range within
$-15.5 < {\rm log}~ SB_{\beta} < -12.5$\footnote{This results
is obtained with the PNs of Table 2, but we eliminated NGC~7017
from this calculation}. 
The lack of
high surface brightness B and BC PNs may simply indicate that they evolve faster
than E and R PNs, similarly to what observed in the Magellanic
Clouds \citep{sea99,sea01}. This is another hint that B and BC PNs may
originate from higher mass progenitors. 

On the other hand, the lack of very low surface brightness
B and BC PNs may indicate that there may be a selection effect 
against low surface brightness B and BC PNs in the Galaxy, possibly due 
to the interstellar extinction which is higher for these PNs, given their
location within the Galactic plane.
The average \hb extinction for each morphological type is given in
column (4) or Table 3. Since on average extinction does anti-correlates
with the distance from the Galactic plane, it roughly indicates that 
interstellar extinction is 
predominant, and the internal extinction is only marginally important in the
overall c$_{\beta}$ values. 

\subsection{Central stars on the Log L -- log T plane, and the 
inferred mass distribution}

Zanstra temperatures of the CSs
were calculated as described by \citet{scs93}. The \heii\
temperatures are the most reliable, as most PNs are optically thick to the 
\heii\ ionizing photons. \hi\ temperatures are generally reliable
for the optically thick PNs. When using \hi~ Zanstra
temperatures we should keep in mind that it is difficult to define which
PN is optically thick and which is thin,
especially in the case of bipolar PNs. Thus the only reliable results are the 
few temperatures calculated from the \heii~ line intensities.

In order to disclose all possible relations
between PNs and their CSs, we have taken a conservative approach.
We first consider the most reliable Zanstra temperature set,
using only \heii\ Zanstra temperatures with 
minimal errors ($\Delta {\rm log} T_{\rm eff} < 0.1$). 
The averages of column (5) in Table 3 refer to this set. 
We do not find marked differences among the average
temperatures of stars hosted by different morphological types.

\citet{cs95} found an average effective temperature
of 142,000 K for stars hosted by bipolar PNs, and 99,000 for those
within
the elliptical (including round) PNs. We do not disclose such a 
difference in the average temperatures.
In Figure 3 we show the location on the log T- log L plane
of the PN CSs, where $\Delta {\rm log} T_{\rm eff} < 0.1$
and $\Delta {\rm log}  L/L_{\odot} < 0.3$ (note that 
the error bars do not account for uncertainties
in the statistical distances).

The HR diagram distribution of Figure 3 shows that stars hosted by
round nebulae are spread on the logT -- logL
plane differently than stars hosted 
by bipolar and bipolar core PNs. Central stars of elliptical PNs 
cover more or less the whole plane, up to the 0.8 M$_{\odot}$ 
evolutionary track. We derive the distribution of the CS's masses
for the different morphologies from Figure 3, and we show it in Figure 4. 
Round nebulae appear to have a peak CS mass 
around 0.55, then the distribution declines for larger masses. 
Central stars of elliptical PNs have a flatter mass distribution. Bipolar  
and bipolar core 
PNs seem to have a flat mass distribution (better compare the round and
elliptical distribution with the sum of bipolar and bipolar core, since these
two later classes are so sparsely populated).

Figure 5 is similar to Figure 3, with the difference that we plot
the HR diagram location of CS PNs whose temperatures have errors 
smaller than 0.4 dex, and where we include the \hi\ Zanstra temperatures.
Figure 6 gives the histograms of the mass
distributions relative to the sample of Figure 5.
Here, the low-mass section of the histogram for elliptical PNs is more
prominent than in Figure 4,
but the overall distributions have the same qualitative
form. These results are indicative of different CS masses for different 
PN shapes. 
The statistical K-S analysis 
applied to this set of temperatures (and the relative luminosities)
does not disclose enormous differences between the effective
temperature and luminosity distributions.

We have analyzed the ratios of \heii\ to \hi\
Zanstra temperatures across morphological types, to have an
indication of their optical thickness. Our aim was to probe the
preliminary results of Figures 1 and 3 in \citet{scs93},
by using the better and more homogeneous sample of Northern Galactic PNs. 
Our results (column [6] of Table 3) shows that,
although there are differences in the average Zanstra ratio for the various
morphologies, 
such differences are milder that previously found
\citep{scs93}. 
We confirm that, on average, bipolar PNs have lower Zanstra
ratio than symmetric PNs, but we can not draw the same conclusion for
bipolar core PNs based on our small sample (see also the higher sigma value).

\subsection{Hints from elemental abundances}

During the evolution of intermediate-mass stars, 
the chemical enrichment of the outer region of the stars
occurs in a series of
dredge-up and nuclear burning events \citep{ir83}, that 
enrich the future PN in different ways, depending on the
stellar mass. By measuring the CNO elemental
abundances we can probe stellar evolution. Carbon depletion and nitrogen enrichment
depend on the progenitor mass, forming a direct connection between
observed progenitor mass (i.e. Population) and chemical content.

The argon and neon abundances in PNs are probes of the
original chemical mix at the time of the formation of the progenitors. 
In fact, these elements are unaffected during the 
evolution of the stars in this mass range. 

A large study of the PN abundances and their correlations to shell
morphology is in preparation \citep{mea02}, based on 
abundances of Galactic PNs newly calculated with a selection
of the available spectral data in the literature.
In Table 4 we show the 
averages of the derived abundances
(all abundances are with respect to hydrogen, by number), calculated
for each morphological type. Note that
we use only reliable elemental abundances (see \citet{mea02}).
With each 
entry, in parenthesis, we report the statistical sample used for that
particular diagnostic. 

Among the significant results, we find that helium abundances are,
on average,
higher in bipolar PNs than in round and elliptical PNs.
The average helium abundance trace nicely the increasing asymmetry,
from round to bipolar PNs. If we consider R+E PNs and B+BC
PNs, the average helium abundance of the two groups differ, quite
significantly, at the 25 
percent level. 

The oxygen abundance averages do not change as much from one group to the
next, with the exception that the average oxygen abundance for bipolar PNs is 
slightly higher than for the other groups.

Bipolar PNs have 
much higher average nitrogen abundance than the other morphological types, 
confirming the results by \citet{pei78} and others. Bipolar core PNs
are mildly nitrogen enriched, based on a rather small PN sample. 
If we compare the R+E
and the B+BC groups, we obtain that the average nitrogen abundances in 
the two groups differ at the 60 percent level. This is a clear signature
that bipolar PNs have massive progenitors.

Other diagnostics do not show remarkable 
differences, or are statistically insignificant. We do not see the
trend found in LMC, where B PNs have higher neon and argon abundances
\citep{sea00}. 

While the abundances used here are probably the best collection available
to date, it is
very difficult to draw sound conclusions on the
basis of sparse spectroscopic data with slits collecting variable
areas of PNs. This is particularly debilitating for multiple shell or
very bipolar PNs, as their abundance gradients can be substantial.
It is worth remarking that, in addition to the above difficulties, 
the abundances are also contaminated by the ISM material,
ingested by the PN during their evolution \citep{vea02}.

\subsection{Evolutionary paths}

As seen in the two preceding sections, there is an indication of mass segregation
among the CSs hosted by symmetric and asymmetric PNs. This has bearings 
on the possible evolutionary paths that lead to the different morphologies. 
The formation of asymmetric PNs has been ascribed to binary evolution, either 
through common envelope (CE) or via wide binary or planetary companions
\citep{sok97},
as well as to the presence of stellar rotation and magnetic fields
\citep{gs99}. 
\citet{sok97} selected a large sample of PNs, and, based solely on their morphology, 
determines the likeliness that these morphologies originate (1) from the
evolution of a single star, (2) from binary evolution, where the member
of the binary system go through the Common Envelope (CE) phase, 
(3) from binary evolution of a pair that avoids the CE phase, or (4)
from the evolution of a star and a planet companion.
We have checked Soker's determinations of the stellar origins of PNs
against our morphological sample. We found, based on our classification scheme and on his
evolutionary scheme, that (1)
no B or BC PNs would originate from single stars; (2) no round PNs progenitors
would have gone through the CE phase;  and (3) no R nor E PNs would
originate through binary
stars, without going through the CE phase. 
If (1) was true, then why the CS masses of
B and BC PNs are distributed differently than those of R and E PNs?
Why B and BC PNs would be found at lower scale heights on the Galactic
plane, and would they have higher helium and nitrogen content? 
Clearly, a theoretical scenario that explains all the observations is
not available at this time. 
In particular, while the binary star scenario can be streched
to explain the Galactic distribution of bipolar PNs \citep{sok1998},
much more stellar evolution modeling is needed to compare
the chemical yields in the cases of single and binary star evolution.

\section{Summary and discussion}

The analysis of Northern Galactic PNs have produced interesting results.
First, there is a 
general indication that bipolar and bipolar core PNs belong to the
same physical class, even if they may be at different evolutionary stages.
This results from the apparent and physical spatial distribution of
B and BC PNs, from the excinction analysis, and from the CSs distributions. The BC sample is barely
large enough for accurate statistical analysis, thus this result
is still to be confirmed in the future. It agrees, though, with the 
analysis of Magellanic Cloud PNs, as discussed by \citet{sea99,sea00,sea01}. 

Second, we have analyzed the CSs via Zanstra analysis. The results
are in moderate agreement with those of \citet{scs93}, after correcting 
for the different morphological classification.
Although the mass histograms of the CSs of Figures 4 and 6 
tell us that bipolar (and bipolar core) PNs have a larger fraction
of massive CSs
(i.e., massive progenitors) than round and elliptical PNs, 
the result is not as sharp as previous results have shown.
Our approach is accurate, yet it lacks true 
luminosity error bars, as the distances to the
nebulae were derived in a  statistical way. Furthermore, the Zanstra analysis
implies the assumption that the CS radiates as a blackbody. 
It is worth noting that, to date, no CS mass has been derived without
strong assumptions on the distances, on the ionized masses, or on model-dependent
parameters. Our conclusion is that, to date, there is not a reliable
way to measure the mass of Galactic PNs or their stellar progenitors, although our
analysis seems to indicate 
that asymmetric PNs were formed from more massive progenitors, on average.

Third, we have investigated the progenitors' populations of the different
morphological PN types: higher mass progenitors should have different chemistry
both from the evolutionary dredge-ups and from the different 
chemistry at star formation. 
We found that the helium and nitrogen (and very marginally oxygen) 
average abundances in bipolar and bipolar core PNs are larger than the corresponding 
values in round and elliptical PNs. In agreement with the theory of stellar evolution,
we confirm that B and BC PNs have more massive progenitors than E and R PNs.

On the other hand, the average neon and argon
abundances are almost identical in round, elliptical, and bipolar
PNs (bipolar core PNs do show high neon and low argon abundances,
but the sample size is very small for statistical significance).
This result is in contrast to a study of
LMC PNs \citep{sea00}, where bipolar (and bipolar core) PNs have higher
neon and argon abundances than round and elliptical PNs. 
We know that there is a larger fraction of
bipolar PNs in the LMC than in the Galaxy \citep{sea01}, and that Galactic
bipolar PNs tend to be located in the Galactic plane, where
the interstellar extinction is  maximum. 
It seems then reasonable that the observed Galactic bipolar PNs are
local, so
their neon and argon abundances reflect the local environment at formation, 
rather than the average Galactic disk environment. 

Last, we conclude that the study of the statistical 
correlations between CSs and PN morphology
has probably reached, in this form, its exploitation.
By increasing the statistical significance and the quality
of the analyzed sample we do
not get sharper results on the statistical 
correlations between CSs and PNs. The future of this type
of studies is, in our view, two-fold. On one hand, the acquisition and interpretation
of CS spectra, and at the same time, the development and improvement of
non-LTE stellar models, is essential to acquire the necessary stellar
data for a non-biased analysis. On the other hand, better nebular analysis
on the individual objects will improve the knowledge of reddening to the
individual PNs, and their global elemental abundances, possibly
extending the analysis to UV spectroscopy, i.e., to reliable
carbon abundances. High
resolution spectroscopy on the PNs will also be essential to confirm the 
morphology of some uncertain cases, where the {\it spatial dimension} 
of the velocity is essential.

A continuing study of those extra-Galactic PNs whose morphologies
are detectable with today's technology (e.g., the LMC, the SMC, and the nearby
dwarf galaxies observed with {\it HST}) will have the double benefit of 
minimizing the ill-effects of the unknown Galactic PN distance scale, and those
of the poorly known reddening, and of allowing 
us to extend and integrate our knowledge of the final phases of 
intermediate-mass stars to
Galaxies of different types and metallicities.
The emerging picture is that PNs generated by more massive progenitors, those
that have been ejected by disk population stars, seem to be bipolar
in shape, while smaller mass stars give birth to round PNs. Elliptical 
PNs are produced by stars in a wider mass range. This result 
is in agreement with the findings of the PN distribution within the 
Galaxy, and with the relative N/H and
He/H abundance of the Galactic PNs sample. It
bears on the 
formation of PNs of different morphological types, and it agrees with 
bipolar PNs to be formed by the evolution of
larger mass stars than round and elliptical PNs. 
While both single and binary star evolutionary models can account for
bipolar PN formation, via common envelope evolution, magnetic fields, and fast
stellar rotation, the binary scenario alone would not comply with our results.

\acknowledgements
It is a pleasure to thank Dick Shaw, Bruce Balick, Stacy Palen, Chris Blades,
and Max Mutchler for participating in our collective PN morphology
classification experiment.
Mark Dickinson is thanked for his 
help in the statistical analysis.
\clearpage

\begin{deluxetable}{ll}
\tablenum{1}
\tablewidth{30pc}
\tablecaption{References}
\tablehead{
\multicolumn{1}{c}{\bf\it Physical parameter} &
\multicolumn{1}{l}{\bf\it Reference}}

\startdata

Stellar magnitude & \citet{abe66}              \\
                   & \citet{cea90}                 \\
                   & \citet{dhw80}       \\
                   & \citet{gp88}         \\
                   & \citet{hw87}  \\
                   & \citet{hh90}              \\
                   & \citet{jk89}            \\
                   & \citet{kal83}               \\
                   & \citet{kf85}           \\
                   & \citet{ksk90}   \\
                   & \citet{koh79}        \\
                   & \citet{kp98}\\
                   & \citet{kea68}          \\
                   & \citet{kjl98}   \\
                   & \citet{lea88}     \\
                   & \citet{lut77}               \\
                   & \citet{sk85}           \\
                   & \citet{tea90}            \\
                   & Tylenda et al. (1991)                                \\

F(6~cm)             &      \citet{aea92}                        \\
                   &  \citet{cks92}                                    \\
                   &  \citet{pz94}             \\
                   &  \citet{wen95}                             \\
              
log($F_{H\beta}$)  &    \citet{aea92}                          \\ 
                   &  \citet{all94}                              \\ 
                   &     \citet{cks92}               \\ 
                   &  \citet{cak96}       \\
Extinction constant &   \citet{all94}                               \\ 
                   &     \citet{cak96}         \\
                   &        \citet{cks92}                                 \\
                   &  \citet{dmc99}         \\
                   &  \citet{sea92}                       \\
                   &  \citet{tea92} 
\enddata
\end{deluxetable}

\clearpage

\begin{deluxetable}{llllll}
\tablenum{2}
\tabletypesize{\scriptsize}

\tablewidth{0pt}
\tablecaption{Morphology, temperatures, and luminosities}
\tablehead{
\colhead{\bf name} &
\colhead{\bf morph.} &
\colhead{\bf log T$_{\rm eff}$} &
\colhead{\bf $\Delta_{\rm log T}$} &
\colhead{\bf log L/L$_{\odot}$} &
\colhead{\bf $\Delta_{\rm log L}$}  \\
}
\startdata
     A4$^{\rm a}$  &  R  &  5.114  & 0.302  & 1.717  &  0.308   \\
        A8$^{\rm a}$  &  R    & 5.187  & 0.301  & 1.932  &  0.302   \\
             A12$^{\rm a,b}$ &  R   & 5.214  & 0.301  & 2.497  &  0.302   \\
       A16  &  R    & 4.936  & 0.040  & 1.800  &  0.142   \\
       A20$^{\rm b}$ &  R    & 5.012  & 0.033  & 2.744  &  0.105   \\
       A28  &  R   & 4.967  & 0.301  & 1.269  &  0.305   \\
       A30  &  R    & 4.874  & 0.013  & 2.600  &  0.068   \\
       A33  &  R    & 4.979  & 0.022  & 2.257  &  0.063   \\
       A39  &  R  & 4.953  & 0.021  & 1.934  &  0.078   \\
       A50$^{\rm a}$  &  R    & 5.178  & 0.301  & 2.027  &  0.301   \\
       A53$^{\rm a,b}$ &  R    & 5.215  & 0.306  & 2.356  &  0.333   \\
       A71  &  R    & 5.113  & 0.015  & 1.830  &  0.042   \\
       A83$^{\rm a}$  &  R    & 4.386  & 0.211  & 2.082  &  3.513   \\
     BV5-3  &  R    & 4.925  & 0.026  & 2.447  &  0.113   \\
       Ba1  &  R    & 5.058  & 0.008  & 2.126  &  0.025   \\
     Cn3-1$^{\rm b}$ &  R    & 4.809  & 0.070  & 3.893  &  0.281   \\
     H3-29$^{\rm a}$  &  R   & 5.954  & 0.301  & 2.496  &  0.301   \\
     H3-75$^{\rm a}$  &  R   & 5.099  & 0.301  & 1.987  &  0.301   \\
     He1-4$^{\rm c}$  &  R    & 5.133  & 0.053  & 3.300  &  0.106   \\
     He1-5$^{\rm a}$  &  R    & 4.894  & 0.301  & 0.000  &      \nodata   \\
   He2-432$^{\rm a}$  &  R  & 4.305  & 0.190  & 4.713  &  3.159   \\
    IC~1454$^{\rm c,d}$  &  R   & 5.067  & 0.301  & 1.760  &  0.304   \\
    IC~4593$^{\rm d}$  &  R    & 4.450  & 0.014  & 3.472  &  0.041   \\
       K1-7$^{\rm a}$  &  R   & 5.038  & 0.301  & 2.100  &  0.301   \\
      K1-9$^{\rm a}$  &  R   & 5.194  & 0.301  & 1.295  &  0.302   \\
         K1-14  &  R    & 4.823  & 0.021  & 3.351  &  0.117   \\
     K1-20$^{\rm a}$  &  R   & 5.011  & 0.301  & 1.349  &  0.301   \\
      K3-2$^{\rm a}$  &  R   & 4.267  & 0.178  & 4.790  &  2.953   \\
     K3-27$^{\rm a}$  &  R   & 5.494  & 0.301  & 2.499  &  0.303   \\
     K3-51$^{\rm a}$  &  R   & 5.534  & 0.301  & 2.946  &  0.303   \\
     K3-56$^{\rm a}$  &  R    & 5.449  & 0.301  & 0.000  &      \nodata   \\
      K3-7$^{\rm a}$  &  R    & 4.179  & 0.145  & 1.931  &  2.365   \\
     K3-73  &  R   & 5.049  & 0.008  & 1.779  &  0.022   \\
     K3-81  &  R    & 4.814  & 0.012  & 3.641  &  0.069   \\
     KjPn1$^{\rm a}$  &  R    & 5.156  & 0.301  & 2.055  &  0.301   \\
      LSA1$^{\rm a}$  &  R   & 4.316  & 0.193  & 2.926  &  3.212   \\
     M1-77$^{\rm d}$  &  R    & 4.320  & 0.006  & 4.430  &  0.030   \\
     M1-80$^{\rm a}$  &  R  & 5.226  & 0.301  & 2.748  &  0.302   \\
     M4-18$^{\rm d}$  &  R    & 4.432  & 0.010  & 3.576  &  0.034   \\
   NGC~2242  &  R    & 5.025  & 0.033  & 2.778  &  0.100   \\
   NGC~2438  &  R   & 5.175  & 0.082  & 2.055  &  0.134   \\
   NGC~3587  &  R    & 5.049  & 0.018  & 1.606  &  0.053   \\
   NGC~6852  &  R   & 5.130  & 0.014  & 2.446  &  0.035   \\
   NGC~6879  &  R  & 4.835  & 0.026  & 3.617  &  0.130   \\
   NGC~6884  &  R    & 4.902  & 0.057  & 3.273  &  0.226   \\
   NGC~6894$^{\rm a}$  &  R    & 4.991  & 0.302  & 2.444  &  0.306   \\
   NGC~7094  &  R   & 4.874  & 0.009  & 3.058  &  0.042   \\
   NGG6842  &  R   & 4.994  & 0.031  & 2.782  &  0.105   \\
       Na1  &  R    & 4.899  & 0.025  & 3.597  &  0.118   \\
     Sa4-1$^{\rm d}$  &  R    & 4.315  & 0.009  & 3.113  &  0.044   \\
       Sd1$^{\rm a}$  &  R    & 4.301  & 0.189  & 2.084  &  3.137   \\
     St3-1$^{\rm a}$  &  R    & 5.134  & 0.301  & 2.144  &  0.301   \\
     Vy2-3  &  R   & 4.764  & 0.013  & 4.153  &  0.047   \\
     We1-5  &  R    & 4.945  & 0.011  & 2.904  &  0.042   \\
   
           A2  &  E  &  5.124  & 0.030  & 2.431  &  0.081   \\
        A3$^{\rm a}$  &  E   & 5.377  & 0.304  & 2.306  &  0.322  \\
         A13  &  E    & 5.174  & 0.110  & 2.759  &  0.181   \\
        A18  &  E    & 5.050  & 0.019  & 1.818  &  0.054   \\
        A19$^{\rm a}$  &  E   & 4.310  & 0.191  & 1.330  &  3.181   \\
        A21$^{\rm d}$  &  E  & 5.062  & 0.079  & 0.932  &  0.045   \\
        A43  &  E    & 4.833  & 0.006  & 2.778  &  0.032   \\
        A46  &  E  & 4.859  & 0.079  & 2.169  &  0.268   \\
        A49$^{\rm a}$  &  E  & 4.303  & 0.190  & 2.393  &  3.148   \\
        A54$^{\rm a}$  &  E   & 4.305  & 0.190  & 2.288  &  3.159   \\
        A57$^{\rm d}$  &  E  & 4.667  & 0.022  & 1.666  &  0.022   \\
        A63$^{\rm d}$  &  E   & 4.623  & 0.251  & 1.922  &  4.084   \\
        A70$^{\rm a}$  &  E   & 5.242  & 0.302  & 1.160  &  0.306   \\
        A72  &  E    & 4.972  & 0.013  & 2.517  &  0.044   \\
        A75$^{\rm a}$  &  E    & 5.463  & 0.301  & 2.580  &  0.304   \\
        A77  &  E   & 4.815  & 0.064  & 3.116  &  0.262   \\
        A78  &  E    & 4.837  & 0.015  & 3.205  &  0.051   \\
        A84  &  E   & 5.029  & 0.061  & 1.692  &  0.146   \\
    Anom20H$^{\rm a}$  &  E    & 4.360  & 0.205  & 1.903  &  3.410   \\
    BD+3036$^{b,c}$  &  E    & 4.705  & 0.113  & 3.133  &  0.090   \\
     CRL618$^{\rm a}$  &  E   & 4.394  & 0.213  & 3.921  &  3.546   \\
        Dd1$^{\rm a}$  &  E    & 5.189  & 0.301  & 1.295  &  0.302   \\
      DeHt2$^{\rm d}$  &  E    & 4.369  & 0.011  & 1.831  &  0.044   \\
       EGB4$^{\rm d}$  &  E   & 4.616  & 0.023  & 2.278  &  0.031   \\
     HaTr14$^{\rm a}$  &  E    & 4.225  & 0.164  & 2.473  &  2.695   \\
      He1-2$^{\rm d}$  &  E    & 4.475  & 0.038  & 3.867  &  0.097   \\
    He2-429$^{\rm a}$  &  E    & 4.362  & 0.205  & 3.771  &  3.418   \\
    He2-430$^{\rm a}$  &  E   & 4.378  & 0.209  & 4.054  &  3.485   \\
    He2-447$^{\rm a}$  &  E    & 5.031  & 0.301  & 2.773  &  0.301   \\
      Hu1-1  &  E    & 5.055  & 0.066  & 2.686  &  0.176   \\
     IC~1295  &  E   & 4.993  & 0.301  & 1.817  &  0.301   \\
     IC~2003  &  E   & 4.954  & 0.028  & 3.445  &  0.109   \\
     IC~2149$^{\rm c,d}$  &  E    & 4.483  & 0.008  & 3.014  &  0.023   \\
      IC~351  &  E   & 5.127  & 0.114  & 3.066  &  0.218   \\
       J320$^{\rm c}$  &  E    & 4.772  & 0.012  & 3.700  &  0.074   \\
        Jn1  &  E   & 4.980  & 0.301  & 1.844  &  0.301   \\
      JnEr1  &  E    & 5.010  & 0.010  & 1.191  &  0.034   \\
      K1-16  &  E   & 5.690  & 0.264  & 1.786  &  0.298   \\
      K1-17$^{\rm a}$  &  E   & 5.343  & 0.301  & 1.517  &  0.302   \\
       K2-1$^{\rm a}$  &  E    & 5.954  & 0.301  & 0.663  &  0.301   \\
       K2-5$^{\rm d}$  &  E   & 4.813  & 0.034  & 1.611  &  0.015   \\
           K3-5$^{\rm a}$  &  E  & 5.375  & 0.301  & 2.644  &  0.302   \\
           K3-21$^{\rm d}$  &  E  & 4.392  & 0.049  & 2.760  &  0.159   \\
   K3-57$^{\rm a}$  &  E  & 5.441  & 0.301  & 3.264  &  0.302   \\
      K3-61$^{\rm a}$  &  E   & 4.299  & 0.188  & 3.537  &  3.126   \\
      K3-64$^{\rm a}$  &  E   & 4.301  & 0.189  & 2.871  &  3.137   \\
      K3-68$^{\rm a}$  &  E    & 5.476  & 0.301  & 2.436  &  0.302   \\
      K3-76$^{\rm a}$  &  E    & 4.265  & 0.178  & 4.240  &  2.939   \\
      K3-79$^{\rm a}$  &  E   & 4.230  & 0.166  & 3.590  &  2.728   \\
      K3-80$^{\rm a}$  &  E  & 4.272  & 0.180  & 3.510  &  2.980   \\
      K3-82$^{\rm a}$  &  E   & 4.427  & 0.220  & 3.291  &  3.657   \\
      K3-92  &  E  & 4.901  & 0.006  & 2.737  &  0.018   \\
       K4-5$^{\rm a}$  &  E   & 5.329  & 0.301  & 1.703  &  0.302   \\
       M1-2  &  E   & 4.736  & 0.024  & 2.384  &  0.155   \\
       M1-6  &  E   & 4.780  & 0.014  & 3.273  &  0.092   \\
      M1-64$^{\rm a}$  &  E    & 4.500  & 0.234  & 3.015  &  3.862   \\
      M1-66$^{\rm a}$  &  E   & 5.037  & 0.301  & 3.080  &  0.301   \\
      M1-73  &  E   & 4.744  & 0.010  & 3.988  &  0.077   \\
      M1-78$^{\rm a}$  &  E   & 4.375  & 0.208  & 2.851  &  3.471   \\
       M2-2$^{\rm a}$  &  E    & 5.216  & 0.301  & 3.229  &  0.302   \\
      M2-40$^{\rm a}$  &  E   & 4.418  & 0.218  & 3.971  &  3.630   \\
      M2-44$^{\rm a}$  &  E   & 5.307  & 0.301  & 3.191  &  0.302   \\
      M2-45$^{\rm a,b}$ &  E    & 4.279  & 0.182  & 4.081  &  3.018   \\
      M2-50$^{\rm a}$  &  E    & 4.975  & 0.301  & 2.713  &  0.301   \\

       M3-3$^{\rm a}$  &  E   & 5.022  & 0.302  & 2.332  &  0.306   \\
        M4-9$^{\rm a,b}$  &  E    & 4.367  & 0.207  & 2.891  &  3.441   \\
             M4-11$^{\rm a}$  &  E   & 5.359  & 0.301  & 2.153  &  0.302   \\
      Me1-1  &  E   & 4.806  & 0.030  & 3.969  &  0.164   \\
      NGC~40$^{\rm c,d}$  &  E   & 4.476  & 0.017  & 3.219  &  0.044   \\
         NGC~1501  &  E  & 4.923  & 0.020  & 3.386  &  0.083   \\
    NGC~1514$^{\rm c}$  &  E   & 4.711  & 0.018  & 4.178  &  0.062   \\
    NGC~2022$^{\rm c}$  &  E   & 5.075  & 0.019  & 3.452  &  0.049   \\
    NGC~2392$^{\rm c}$  &  E    & 4.870  & 0.007  & 3.828  &  0.034   \\
    NGC~6210$^{\rm c}$  &  E    & 4.814  & 0.014  & 3.477  &  0.086   \\
    NGC~6543$^{\rm c,d}$  &  E   & 4.690  & 0.051  & 3.080  &  0.045   \\
    NGC~6720  &  E   & 5.148  & 0.026  & 2.856  &  0.059   \\
    NGC~6741$^{\rm b}$ &  E   & 5.356  & 0.224  & 2.870  &  0.153   \\
    NGC~6781$^{\rm c}$  &  E  & 5.022  & 0.040  & 2.104  &  0.112   \\
    NGC~6826$^{\rm c,d}$  &  E   & 4.511  & 0.034  & 3.321  &  0.074   \\
    NGC~6853  &  E   & 4.777  & 0.016  & 1.348  &  0.102   \\
    NGC~6905$^{\rm c}$  &  E   & 5.117  & 0.042  & 3.668  &  0.093   \\
    NGC~7008  &  E  & 5.701  & 0.254  & 3.063  &  0.301   \\
    NGC~7048$^{\rm c}$  &  E  & 5.256  & 0.301  & 2.222  &  0.304   \\
    NGC~7354$^{\rm c}$  &  E  & 4.989  & 0.301  & 3.613  &  0.301   \\
    NGC~7662$^{\rm c}$  &  E    & 5.051  & 0.069  & 3.395  &  0.184   \\
        PB9$^{\rm a}$  &  E  & 4.988  & 0.301  & 3.273  &  0.301   \\
       PC19$^{\rm a}$ &  E   & 4.949  & 0.301  & 3.818  &  0.301   \\
     Pe1-15$^{\rm d}$  &  E   & 4.422  & 0.036  & 3.616  &  0.111   \\
     Pe1-16  &  E   & 4.886  & 0.029  & 4.133  &  0.136   \\
     Pe1-20$^{\rm a}$  &  E  & 5.608  & 0.301  & 2.177  &  0.303   \\
     Pe1-21$^{\rm a}$  &  E  & 5.954  & 0.301  & 2.610  &  0.301   \\
        Pu1  &  E  & 5.002  & 0.041  & 1.801  &  0.130   \\
        Pu2$^{\rm d}$  &  E   & 4.606  & 0.062  & 1.371  &  0.086   \\
      Vy1-1  &  E  & 4.730  & 0.006  & 3.687  &  0.041   \\
      Vy1-2  &  E   & 5.075  & 0.084  & 3.275  &  0.285   \\
      WeSb3$^{\rm d}$  &  E   & 4.599  & 0.248  & 1.344  &  4.050   \\

       A79$^{\rm a,c}$  &  B    & 4.389  & 0.212  & 2.404  &  3.527   \\
      HDW5$^{\rm d}$  &  B   & 4.657  & 0.029  & 1.258  &  0.031   \\
   He2-428$^{\rm a}$  &  B    & 4.248  & 0.172  & 2.837  &  2.839   \\
   He2-437$^{\rm a}$  &  B    & 4.378  & 0.209  & 3.313  &  3.485   \\
      K3-1$^{\rm a}$  &  B    & 5.431  & 0.301  & 2.665  &  0.302   \\
      K3-4$^{\rm a}$  &  B    & 4.326  & 0.196  & 2.896  &  3.262   \\
           K3-24$^{\rm a}$  &  B   & 4.207  & 0.157  & 4.221  &  2.569   \\
     K3-34$^{\rm a}$  &  B    & 5.162  & 0.301  & 1.858  &  0.301   \\
     K3-45$^{\rm a}$  &  B    & 4.215  & 0.160  & 3.386  &  2.624   \\
     K3-46$^{\rm a}$  &  B    & 4.344  & 0.201  & 2.211  &  3.344   \\
     K3-58$^{\rm a}$  &  B    & 4.288  & 0.185  & 3.179  &  3.068   \\
     K3-91  &  B   & 5.234  & 0.301  & 1.790  &  0.302   \\
     K3-93$^{\rm a}$  &  B    & 5.113  & 0.301  & 2.242  &  0.301   \\
     K3-94$^{\rm a}$  &  B    & 5.283  & 0.301  & 2.388  &  0.302   \\
     K4-55$^{\rm a}$  &  B   & 4.301  & 0.189  & 3.338  &  3.137   \\
      M1-8  &  B    & 5.206  & 0.301  & 2.115  &  0.302   \\
           M1-57  &  B  &  5.116  & 0.147  & 3.762  &  0.271   \\
     M1-59$^{\rm a}$  &  B    & 5.091  & 0.301  & 0.000  &      \nodata   \\
     M1-75$^{\rm a}$  &  B   & 5.462  & 0.301  & 2.154  &  0.302   \\

     M2-46$^{\rm a}$  &  B   & 4.246  & 0.171  & 3.495  &  2.824   \\
     M2-48$^{\rm a}$  &  B  & 4.301  & 0.189  & 3.591  &  3.137   \\
     M2-52$^{\rm a}$  &  B    & 5.603  & 0.301  & 2.970  &  0.303   \\
     M3-28$^{\rm a}$  &  B    & 4.260  & 0.176  & 3.881  &  2.912   \\
     M3-55$^{\rm a}$  &  B    & 4.464  & 0.227  & 3.312  &  3.769   \\
     M4-14$^{\rm a}$  &  B   & 4.378  & 0.209  & 3.243  &  3.485   \\
     M4-17$^{\rm a}$  &  B   &  5.103  & 0.301  & 2.125  &  0.301   \\
    NGC~650  &  B    & 5.123  & 0.043  & 2.162  &  0.091   \\
      NGC~2346$^{\rm c}$  &  B   & 4.760  & 0.011  & 3.835  &  0.042   \\
   NGC~6778$^{\rm b}$ &  B    & 5.030  & 0.070  & 2.605  &  0.199   \\
   NGC~6881  &  B  & 4.993  & 0.301  & 0.000  &      \nodata   \\
   NGC~7026$^{\rm c}$  &  B   & 4.965  & 0.029  & 2.913  &  0.108   \\
   NGC~7027$^{\rm c}$  &  B    & 5.275  & 0.301  & 5.989  &  0.301   \\
    Pe1-14$^{\rm a}$  &  B  & 4.196  & 0.152  & 4.299  &  2.491   \\
    Pe1-17$^{\rm a}$  &  B    & 5.410  & 0.301  & 2.462  &  0.302   \\
    Sh1-89$^{\rm a}$  &  B    & 5.221  & 0.301  & 1.793  &  0.302   \\
    Sh2-71$^{\rm a}$  &  B    & 4.890  & 0.011  & 3.353  &  0.041   \\

       A80  & BC    & 5.114  & 0.014  & 1.886  &  0.035   \\
       A82$^{\rm c}$  & BC   & 4.863  & 0.031  & 3.071  &  0.135   \\
     He1-1$^{\rm a}$  & BC    & 5.404  & 0.301  & 3.203  &  0.302   \\
     He1-6$^{\rm a}$  & BC   & 5.082  & 0.301  & 2.102  &  0.301   \\
     Hu1-2  & BC   & 5.058  & 0.066  & 3.100  &  0.174   \\
     K3-36  & BC    & 4.890  & 0.025  & 3.229  &  0.120   \\
     K3-63  & BC  & 4.847  & 0.301  & 3.839  &  0.302   \\
     K3-79$^{\rm a}$  & BC    & 4.230  & 0.166  & 3.590  &  2.728   \\
     K3-84$^{\rm a}$  & BC    & 5.000  & 0.301  & 2.606  &  0.301   \\
      M1-7  & BC   & 5.045  & 0.302  & 2.353  &  0.307   \\
     M1-79  & BC   & 5.025  & 0.302  & 2.483  &  0.306   \\
     M2-51$^{\rm c}$  & BC    & 4.776  & 0.028  & 3.635  &  0.153   \\
     M2-53  & BC   & 5.035  & 0.007  & 2.521  &  0.023   \\
     M2-55  & BC   & 5.031  & 0.038  & 1.883  &  0.099   \\
   NGC~6058  & BC    & 4.834  & 0.005  & 3.435  &  0.030   \\
   NGC~6772$^{\rm b}$  & BC   & 5.147  & 0.042  & 2.160  &  0.094   \\
   NGC~6804$^{\rm b}$ & BC   & 4.954  & 0.007  & 3.804  &  0.030   \\
   NGC~7139  & BC    & 5.020  & 0.301  & 1.921  &  0.301   \\
      PB10$^{\rm a}$  & BC   & 5.343  & 0.301  & 3.612  &  0.302   \\

\enddata
\tablenotetext{a}{Temperature, luminosity, and errors for this nebula are from
cross-over analysis. The temperatures should be considered upper limits.}
\tablenotetext{b}{Morphology and dimensions measured from the \citet{scm92} images.}
\tablenotetext{c}{Morphology and dimensions measured from the \citet{bal87} images.}
\tablenotetext{d}{Temperature, luminosity, and errors for this nebula are from H I Zanstra analysis.}

\end{deluxetable}
\clearpage

\begin{deluxetable}{lccccc}
\tablewidth{0pt}
\tablenum{3}
\tablecaption{Physical diagnostics. I. Galactic distribution, extinction, and Zanstra
temperatures}
\tablehead {
\multicolumn{1}{l}{\bf morph.} &
\multicolumn{1}{c}{\bf $<|b|>$} &
\multicolumn{1}{c}{\bf $<|z|>$ } &
\multicolumn{1}{c}{\bf $c_{\beta}$} &
\multicolumn{1}{c}{\bf $< T_{\rm eff}^a >$} &
\multicolumn{1}{c}{\bf $<T_{\rm eff, He II} / T_{\rm eff, H I}>$ } \\

\multicolumn{1}{l} {} &
\multicolumn{1}{c} {[deg.]} &
\multicolumn{1}{c} {[kpc]} &
\multicolumn{1}{c}  {} &
\multicolumn{1}{c} {$10^3$ [K]} &
\multicolumn{1}{c} {} \\
}

\startdata
R& 12.6 [12.6]& 0.73 [0.7]& 0.63 [0.73]&        95 [25]&  1.5 [0.6] \\

E& 7.2 [7.1]& 0.38 [0.48]&      0.95 [0.77]& 87 [27]&  1.4 [0.6] \\

BC& 6.9 [6.9]& 0.37 [0.53]& 1.03 [0.56]&        97 [27]&  1.5 [0.7] \\

B& 2.9 [2.5]& 0.21 [0.24]& 1.39 [0.63] &        94 [29]& 1.2 [0.5] \\

R + E & 8.9 [9.6]& 0.59 [0.5]& 0.84 [0.77] & 90 [26]&  1.43 [0.6] \\

B + BC& 4.1 [6]& 0.21 [0.24]&   1.26 [0.63]&  96 [27]&  1.3 [0.6] \\

\enddata

\tablenotetext{a}{$\Delta {\rm log} T_{\rm eff} < 0.1$, \heii~ Zanstra temperatures only.}

\end{deluxetable}

\clearpage

\begin{deluxetable}{llllll}
\tablenum{4}
\tablewidth{0pt}
\tablecaption{Physical diagnostics. II. Abundances}
\tablehead{
\multicolumn{1}{c}{\bf morph.} &
\multicolumn{1}{l}{\bf $< {\rm He/H} >^a$} &
\multicolumn{1}{l}{\bf $< {\rm O/H} >$} &
\multicolumn{1}{l}{\bf $< {\rm N/H} >$} &
\multicolumn{1}{l}{\bf $< {\rm Ne/H} >$} &
\multicolumn{1}{l}{\bf $< {\rm Ar/H} >$} \\
 }
\startdata

R&  1.10 (18)& 3.61 (16) & 1.71 (14)& 0.96 (16)& 1.29 (7)  \\

E&  1.20 (33)& 3.34 (32)& 1.51 (31)& 0.87 (32)& 1.28 (27) \\ 

BC& 1.39 (9)& 3.22 (8)& 1.92 (6)& 1.36 (5)& 0.60 (4) \\

B& 1.49 (16)& 4.05 (13)& 5.66 (13)& 0.94 (11)& 1.36 (9) \\
\enddata
\tablenotetext{a}{He/H abundances are multiplied by 10; O/H, N/H, and Ne/H 
abundances are multiplied by 10$^4$; Ar/H abundances are multiplied by 10$^6$.}

\end{deluxetable}

\clearpage\clearpage

\begin{figure}
\plotone{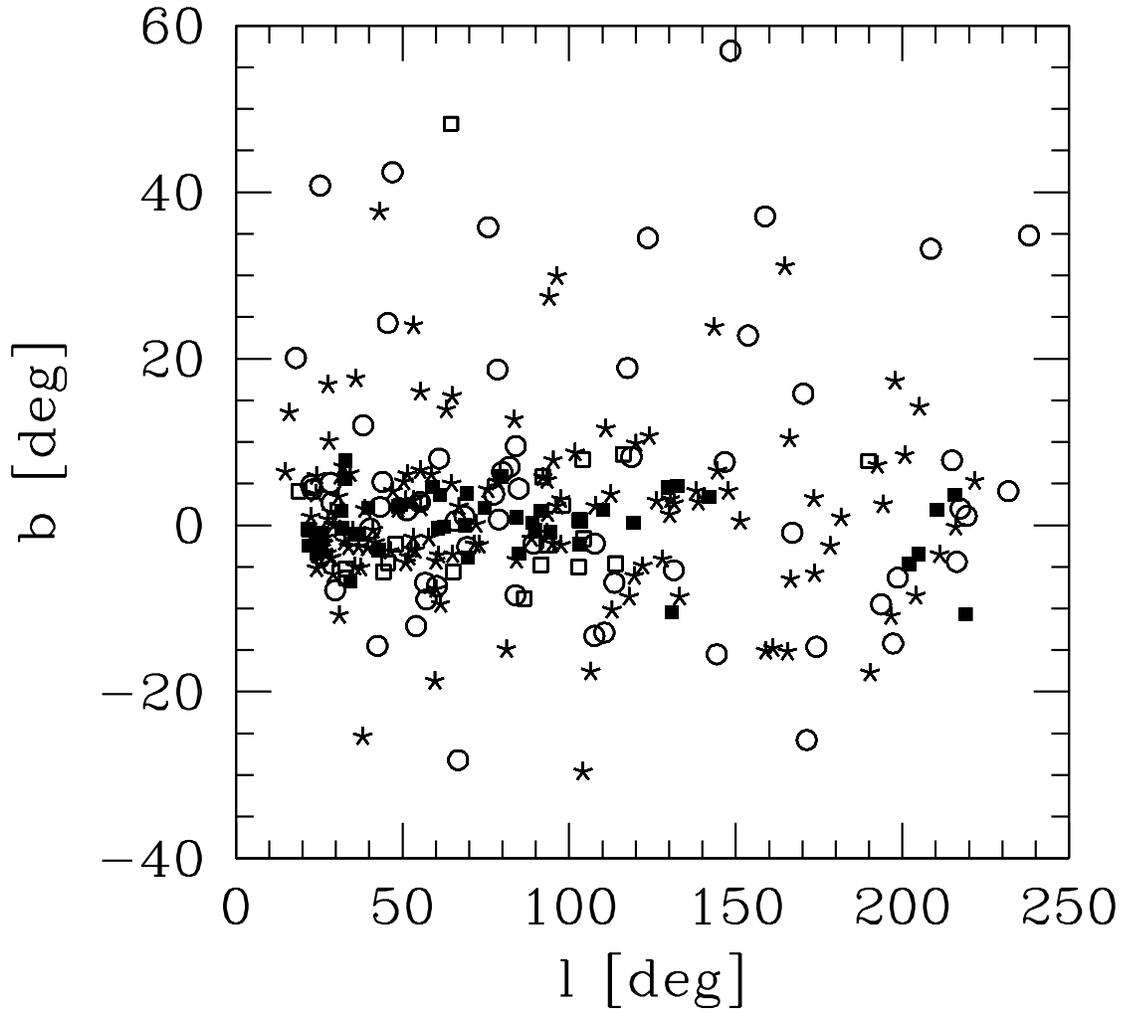} 
\caption{Apparent spatial distribution of the sample PNs
(Galactic longitude versus Galactic latitude). 
Open circles: R PNs; asterisks: E PNs;
filled squares: B PNs; open squares: BC PNs.}
\end{figure}

\begin{figure}
\plotone{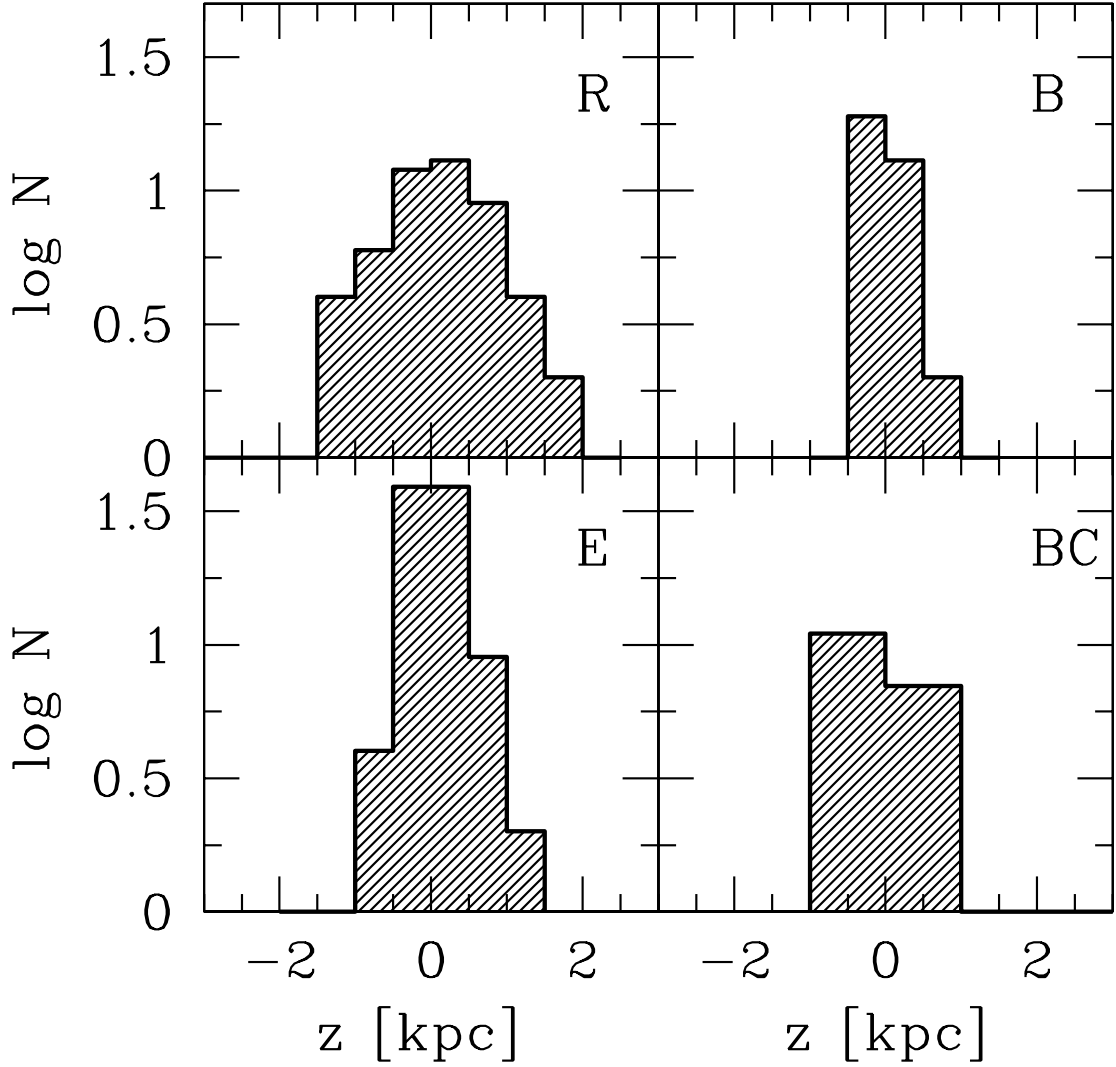} 
\caption{Logarithmic distribution of the scale height for
the different morphological types. Upper left: round PNs; lower left:
upper right: bipolar PNs; lower right: bipolar core PNs. }
\end{figure}

\begin{figure}
\plotone{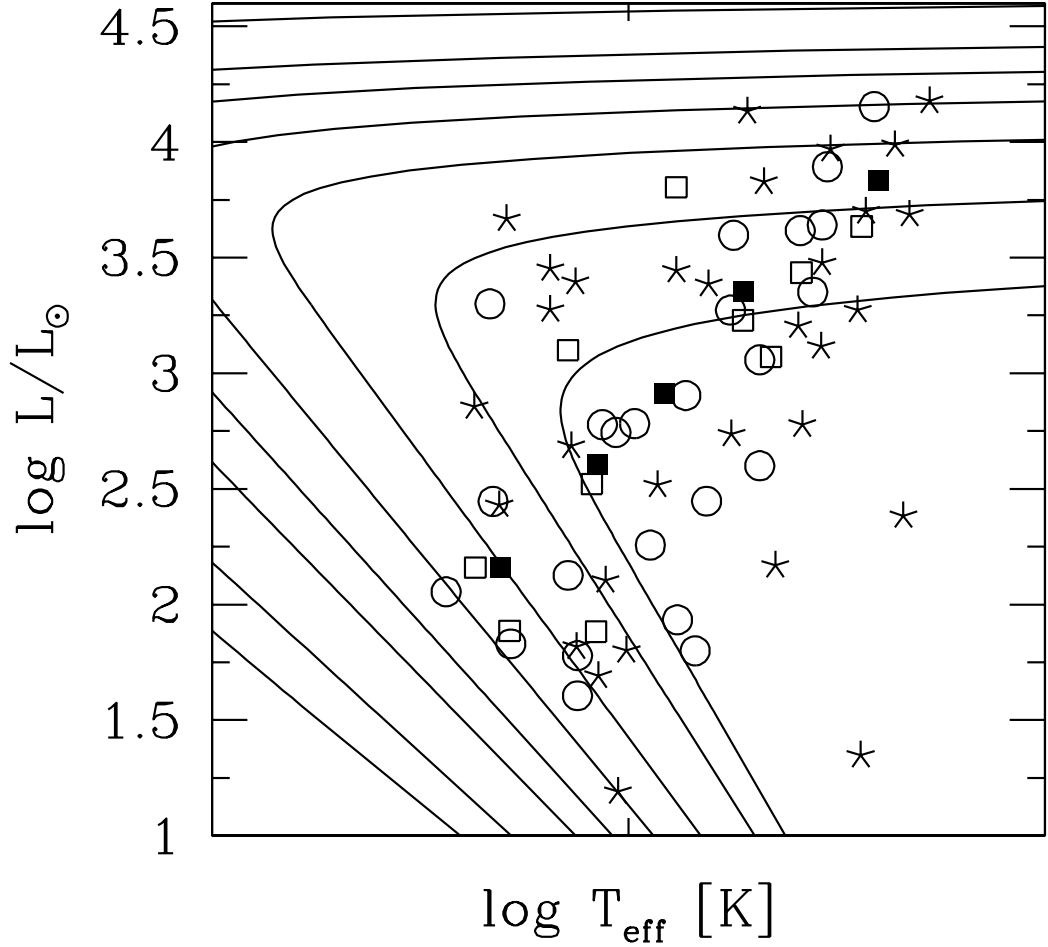} 
\caption{Central stars of Planetary Nebulae in the HR diagram. 
Symbols as in Figure 1.
The constraints in the temperature and luminosity errors, in this
figure, are $\Delta T_{\rm eff}<0.1$ dex and 
$\Delta L< 0.3$ dex. We use the \heii\ Zanstra temperatures only.
Evolutionary tracks are for M=0.55, 0.6, 0.7, 0.8
0.9, 1, 1.2, and 1.4 M$_{\odot}$, from \citet{sr00}.}
\end{figure}

\begin{figure}
\plotone{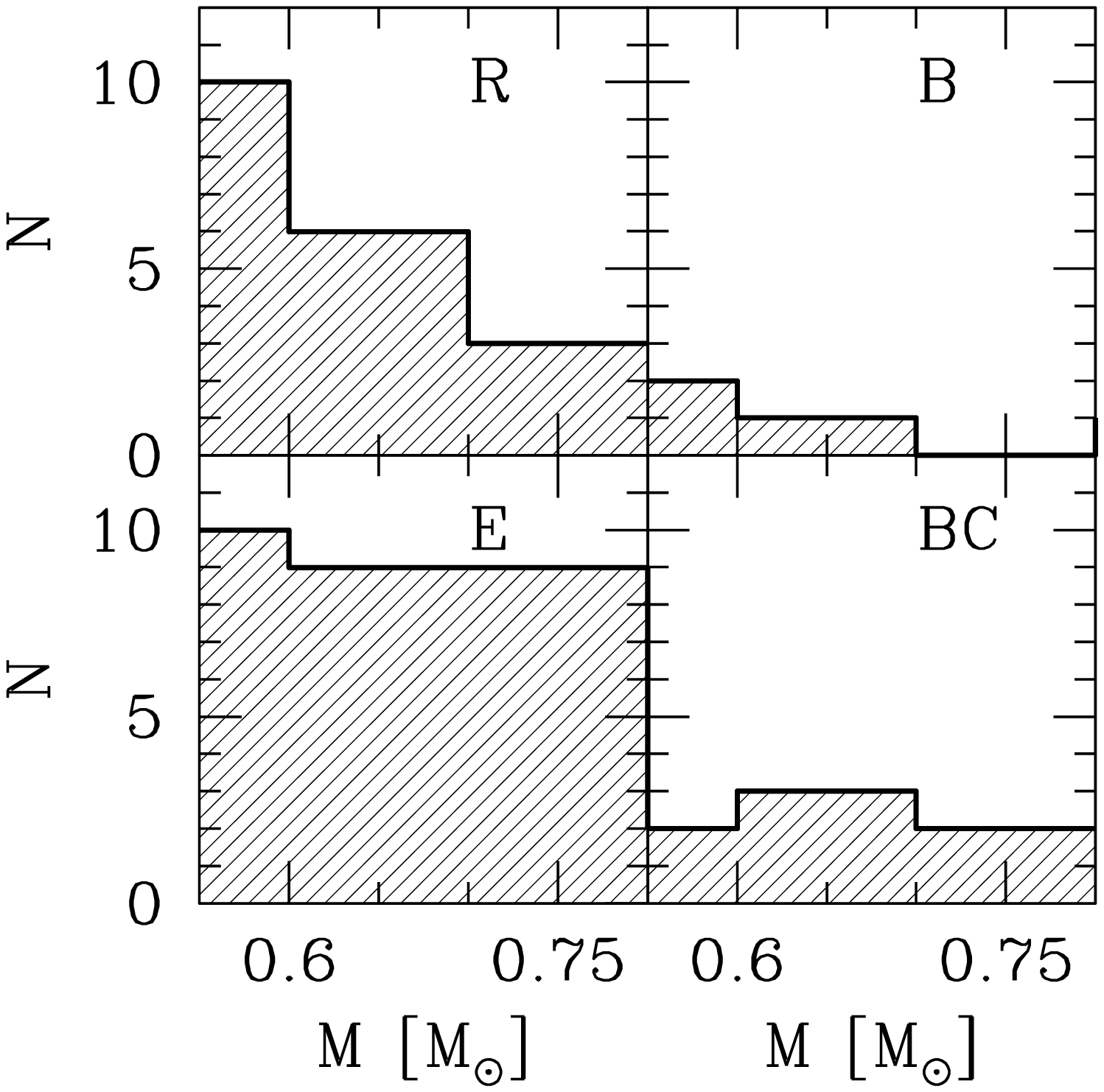} 
\caption{Mass distribution of the CSs, from the sample of
Figure 3. Upper left: CSs of round PNs; lower left:
elliptical PNs; upper right: bipolar PNs; lower right: bipolar core PNs.}
\end{figure}

\begin{figure}
\plotone{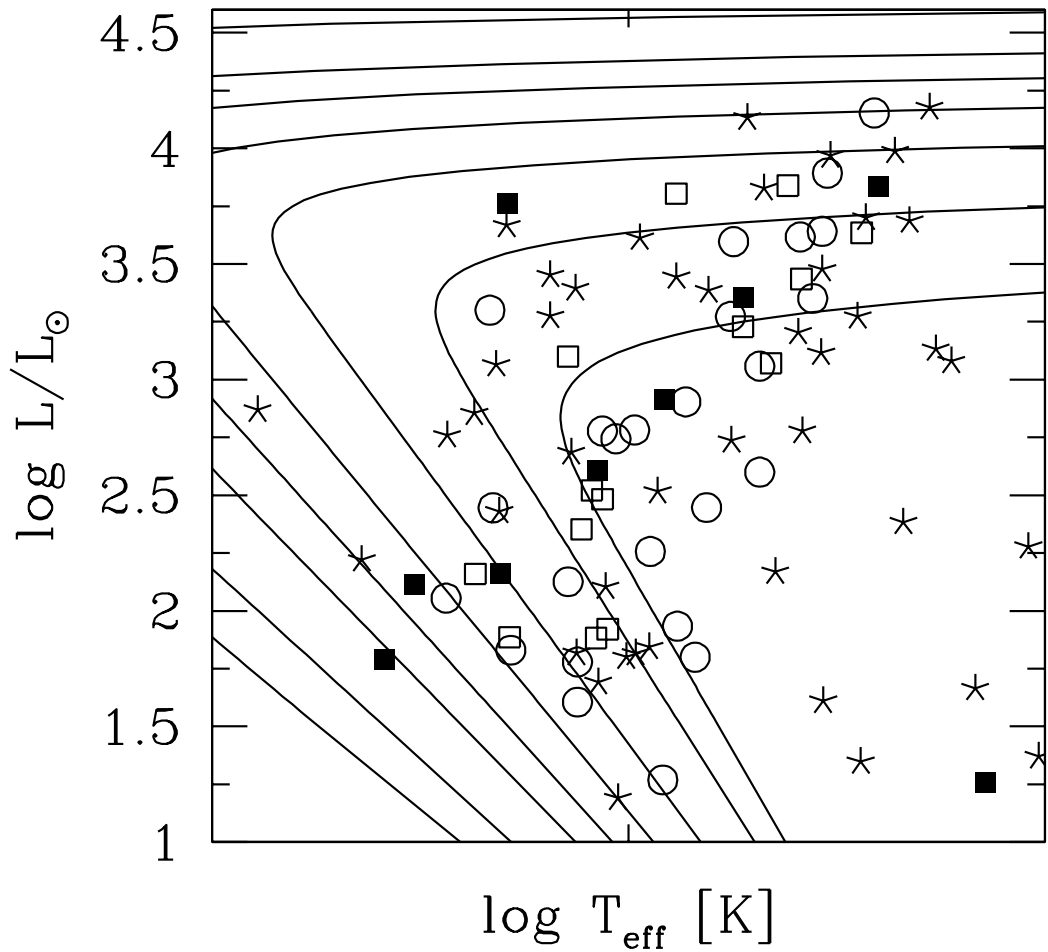} 
\caption{Same plot than in Figure 3, except with 
$\Delta T_{\rm eff}<0.4$ dex and 
$\Delta L< 0.8$ dex. We use the \heii\ Zanstra temperature and, 
when those are not available, the
\hi\ Zantsra temperatures. Symbols and tracks as in Figure 3.}
\end{figure}

\begin{figure}
\plotone{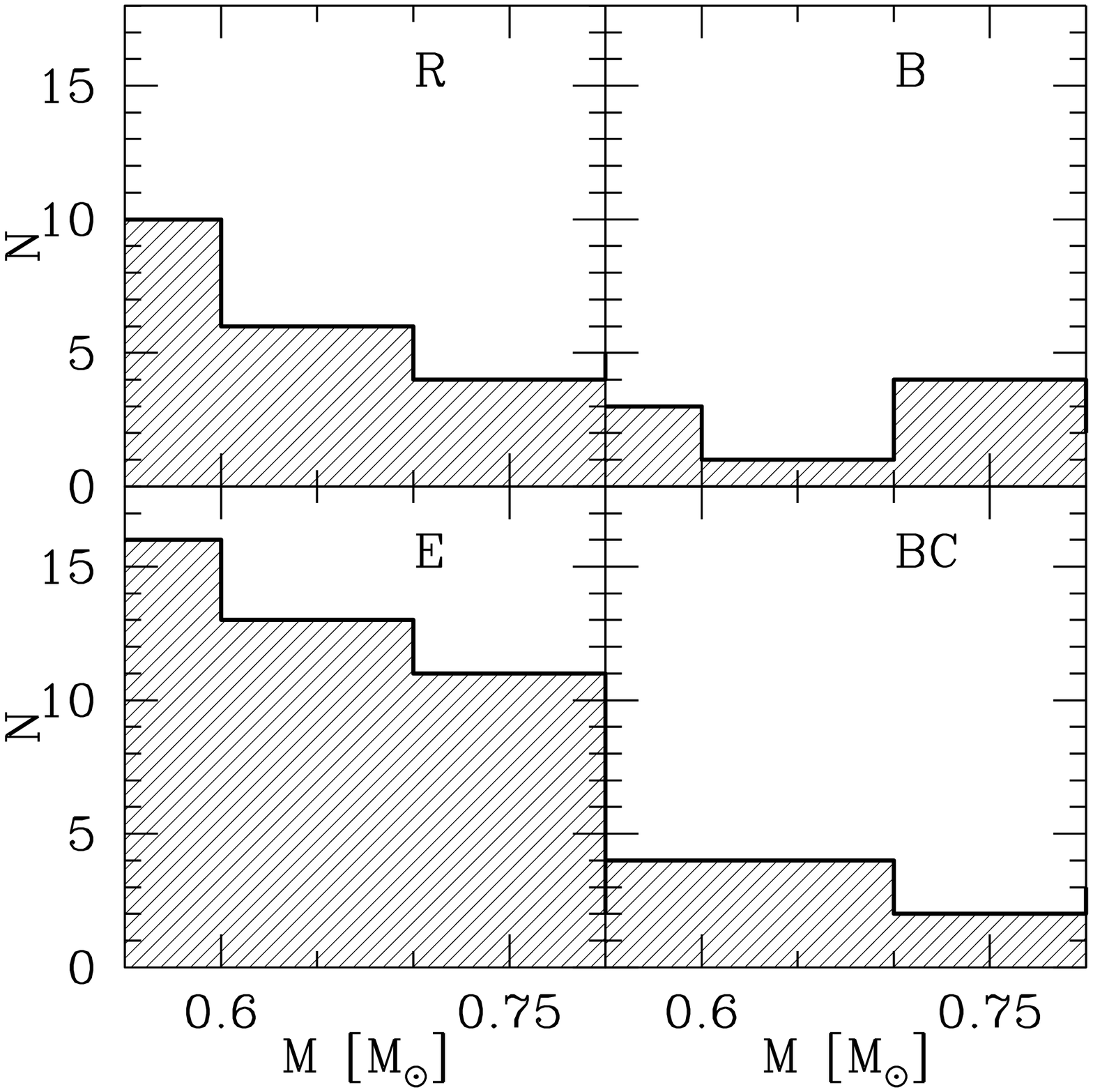} 
\caption{Mass distribution of the CSs, from the sample of
Figure 5. Upper left: CSs of round PNs; lower left:
elliptical PNs; upper right: bipolar PNs; lower right: bipolar core PNs.}
\end{figure}

\end{document}